\documentclass[%
 reprint,
 amsmath,amssymb,
 aps,
 prmaterials,
longbibliography
]{revtex4-2}

\usepackage[utf8x]{inputenc}
\usepackage[english]{babel}
\usepackage{graphicx}
\usepackage{dcolumn}
\usepackage{bm}
\usepackage{siunitx}
\usepackage[version=4]{mhchem}
\usepackage[dvipsnames]{xcolor}

\DeclareMathOperator\erf{erf}

\begin{document}

\preprint{}

\title{Asymmetries in triboelectric charging: generalizing mosaic models to different-material samples and sliding contacts}

\author{Galien Grosjean}
 \email{galien.grosjean@ista.ac.at}
\author{Scott Waitukaitis}%
\affiliation{%
Institute of Science and Technology Austria (ISTA), Lab Building West, Am Campus 1, 3400 Klosterneuburg, Austria
}%

\date{\today}

\begin{abstract}
Nominally identical materials exchange net electric charge during contact through a mechanism that is still debated. `Mosaic models', in which surfaces are presumed to consist of a random patchwork of microscopic donor/acceptor sites, offer an appealing explanation for this phenomenon. However, recent experiments have shown that global differences persist even between same-material samples, which the standard mosaic framework does not account for. Here, we expand the mosaic framework by incorporating global differences in the densities of donor/acceptor sites. We develop an analytical model, backed by numerical simulations, that smoothly connects the global and deterministic charge transfer of different materials to the local and stochastic mosaic picture normally associated with identical materials. Going further, we extend our model to explain the effect of contact asymmetries during sliding, providing a plausible explanation for reversal of charging sign that has been observed experimentally.
\end{abstract}

\maketitle

\section{Introduction}

Contact electrification (CE), also known as triboelectrification or tribocharging, is the exchange of electric charge between objects after touching or rubbing against each other. It occurs in settings as varied as industrial powder flows~\cite{rescaglio_combined_2017}, volcanic plumes~\cite{cimarelli_volcanic_2022}, dust storms~\cite{zhang_reconstructing_2020}, or the early stages of planet formation~\cite{steinpilz_electrical_2020}. Fundamental issues, including the identity of the charge carrier(s) and the property (or properties) driving charge exchange, are still vigorously debated \cite{lacks_long-standing_2019}. In addition to these issues, one particularly confounding aspect of CE is that samples made of the same material systematically exchange charge \cite{shaw_electrical_1926}. This is particularly prevalent in granular systems, where large amount of charge can build up from repeated collisions, leading for example to the electrification of dust storms and thunderclouds \cite{rescaglio_combined_2017,cimarelli_volcanic_2022,zhang_reconstructing_2020,steinpilz_electrical_2020}.

\begin{figure}
  \centering
  \includegraphics[width=8.6cm]{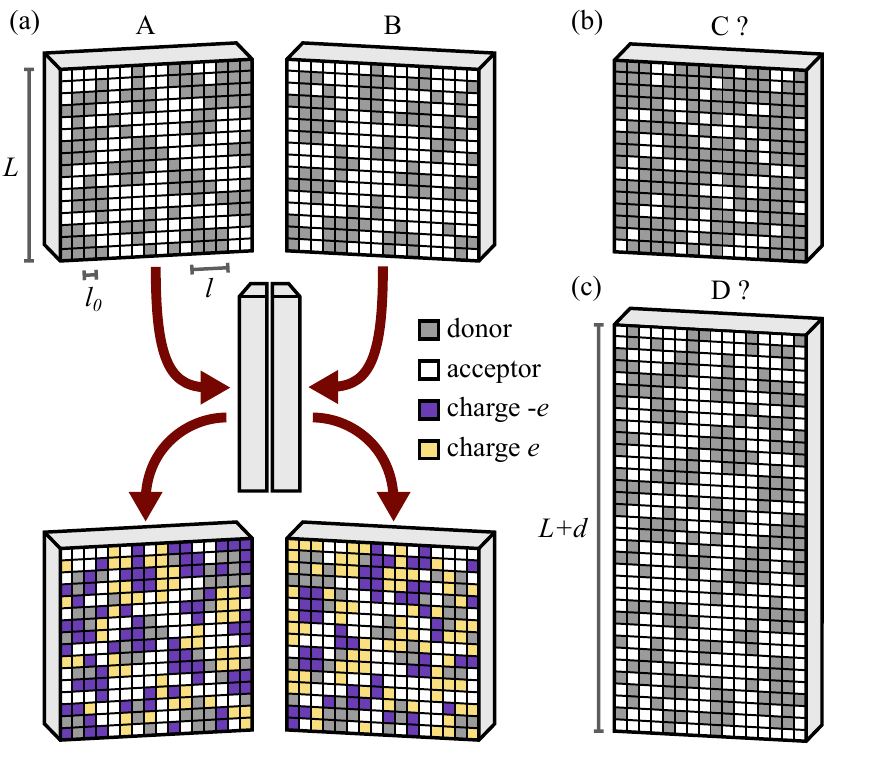}
  \caption{ Mosaic models.  (a) In the standard mosaic formulation, two surfaces, A and B, of size $L \times L$ are thought to consist of donor sites with probability $p$, and acceptors with probability $1-p$. Sites have a size $l_0$ and can be spatially correlated over a length $l$. During contact, donors give a charge $e$ to facing acceptors, with a certain probability $\alpha$. In this paper, we adapt this model in two ways.  (b) First, we study contact between samples of different donor probability, e.g.~grid A \textit{vs.}~C with $p_{\mathrm{A}} \neq p_{\mathrm{C}}$. (c) Second, we study situations were one smaller grid slides a distance $d$ over another longer one, e.g. grid A \textit{vs.}~D.
  }
  \label{fig1} 
\end{figure}

Models for different-material CE and same-material CE are usually considered separately, in large part due to a geometric distinction in their underlying frameworks. Models for different-material CE often consider transfer of charge due to an underlying \textit{global} parameter, \textit{i.e.}~one that is spatially uniform over the surface of a material. For example, one can point to the work function in the case of different metals \cite{lowell_contact_1975}, or Lewis basicity as one candidate (of very many) to explain the case of insulators~\cite{zhang_rationalizing_2019}. On the other hand, models for same-material CE often consider transfer due to an underlying parameter that is \textit{local}, \textit{i.e.}~allowed to vary over the surface of a material. Surfaces can be thought of as consisting of a very large number of microscopic donor and acceptor sites, dividing the surface of the material into a grid or mosaic. During contact, if a donor on one surface touches an acceptor on the other, charge exchange occurs locally, and net charge exchange becomes a stochastic result of all such `micro-exchanges' \cite{apodaca_contact_2010}. In Fig.~\ref{fig1}, we illustrate this process with two square grids, A and B, randomly populated with donor and acceptor sites. The exact nature of the hypothesized, locally-varying charge-driving parameter is unknown, but proposed candidates include adsorbed water islands~\cite{knorr_squeezing_2011}, material from the sample~\cite{baytekin_mosaic_2011}, patterns of strain in the material~\cite{sow_dependence_2012}, anions and cations produced by heterolytic cleavage of polymers~\cite{burgo_triboelectricity_2012,balestrin_triboelectricity_2014}, just to name a few. Regardless of what the parameter is, mosaic models are appealing and widely favored because they permit identical materials to be `identical' in a global sense, but nonetheless still exchange charge \cite{apodaca_contact_2010,xie_contact_2013,yu_numerical_2017,harris_temperature_2019,grosjean_quantitatively_2020}. 

Mosaic models are qualitatively consistent with experimental data obtained with Kelvin Probe Force Microscopy (KPFM) and the Scanning Kelvin Probe method (SKP), which indicate regions of alternating polarity on CE-charged surfaces with lengthscales ranging from nanometers to centimeters \cite{hull_method_1949,terris_contact_1989,shinbrot_spontaneous_2008,knorr_squeezing_2011,baytekin_mosaic_2011,burgo_triboelectricity_2012,barnes_heterogeneity_2016}. At the same time, there is ample room for interpretation. For example, converting KPFM and SKP voltage data to charge data is non-trivial \cite{pertl_quantifying_2022}, and experiments to date have not been able to connect the fluctuating features seen at smaller local scales to the net charge transferred at the global scale. Recent results by Sobolev \textit{et al.}~suggest that much of the mosaic patterns observed could be explained by discharge that occurs when samples are separated \textit{after} CE \cite{sobolev_charge_2022}, and experiments that look at charge transfer from surface-exploring contacts on granular particles show features reminiscent of \textit{both} global and local mechanisms~\cite{grosjean_single-collision_2023}.

In this paper, we build a bridge between the local models of same-material CE and the global models of different material CE by extending mosaic models to account for asymmetries in the densities of donors/acceptors.  For instance, we can illustrate this by considering the grids A and C in Fig.~\ref{fig1}, where C has a much higher density of donors than A. We reveal a continuous transition from a `same-material' regime where charge exchange is stochastic to a `different material' regime where it is deterministic. We validate our analytical treatment of the problem with numerical simulations that mimic the essential ingredients of the hypothesized charge transfer process. We illustrate the value of these results by showing they can qualitatively explain real world situations, \textit{e.g.}~our own experiments with surface-exploring granular contacts~\cite{grosjean_single-collision_2023}.

Going further, we also show that introducing donor/acceptor asymmetries in mosaic models allows us to make sense of older experimental results regarding polarity reversal during sliding~\cite{shaw_tribo-electricity_1928}. Sliding contacts~\cite{shaw_electrical_1926, shaw_tribo-electricity_1928, lowell_triboelectrification_1986-1, lowell_triboelectrification_1986} and, to some extent, rolling contacts~\cite{grzybowski_dynamic_2003,wiles_effects_2004,thomas_iii_patterns_2008} are situations where the contact itself introduces an asymmetry. Contact is spread over a much larger area on the `rubbed' sample than on the `rubber', \textit{e.g.}~grids A and D in Fig.~\ref{fig1}. Using our generalization of mosaic models, we show that charging can be non-monotonous with sliding distance, granted the initial donor/acceptor probabilities differ. In some cases, this can lead the net charge of the `rubber' to flip sign. Both non-monotonous charging and sign flips have been observed experimentally, but were previously attributed to contaminants on the surface~\cite{shaw_tribo-electricity_1928, lowell_triboelectrification_1986}. The value of our results here, therefore, is to offer an alternative explanation to this phenomenon which naturally emerges from donor/acceptor asymmetries.

\section{Accounting for donor/acceptor asymmetry}

\begin{table*}
\begin{tabular}{|c|c|c|c|} \hline
\textbf{Outcome} & \textbf{1. Strict probabilities} & \textbf{2. Relaxed probabilities} & \textbf{3. Approximate relaxed probabilities} \\ \hline
transfer A$\rightarrow$B & $p_{\mathrm{A}}(1-p_{\mathrm{B}})\,\alpha$ & $p_{\mathrm{A}}(1-p_{\mathrm{B}})\,\alpha\left(1-p_{\mathrm{B}}(1-p_{\mathrm{A}})\,\alpha\right)$ & $\sim p_{\mathrm{A}}(1-p_{\mathrm{B}})\,\alpha$\\ \hline
transfer A$\leftarrow$B & $p_{\mathrm{B}}(1-p_{\mathrm{A}})\,\alpha$ & $p_{\mathrm{B}}(1-p_{\mathrm{A}})\,\alpha\left(1-p_{\mathrm{A}}(1-p_{\mathrm{B}})\,\alpha\right)$ & $\sim p_{\mathrm{B}}(1-p_{\mathrm{A}})\,\alpha$ \\ \hline
transfer A$\rightleftarrows$B & 0 & $p_{\mathrm{A}}p_{\mathrm{B}}(1-p_{\mathrm{B}})(1-p_{\mathrm{B}})\,\alpha^2$ & $\sim0$ \\ \hline 
no transfer & $1-(p_{\mathrm{A}}+p_{\mathrm{B}}-2p_{\mathrm{A}}p_{\mathrm{B}})\,\alpha$ & $\left(1-p_{\mathrm{A}}(1-p_{\mathrm{B}})\right)\left(1-p_{\mathrm{B}}(1-p_{\mathrm{A}})\right)\alpha^2$ & $\sim1-(p_{\mathrm{A}}+p_{\mathrm{B}}-2p_{\mathrm{A}}p_{\mathrm{B}})\,\alpha$  \\ \hline
\end{tabular}
\caption{Transfer probabilities can be considered in three different ways. Strictly, there are three possible outcomes for each contacting pair of sites: A$\rightarrow$B transfer, A$\leftarrow$B transfer, or no transfer. The probabilities of each outcome are presented in column 1. Column 2 presents the relaxed probabilities, where we assume that A$\rightarrow$B and A$\leftarrow$B transfers are independent. This means that the probability of simultaneous transfer in both directions is nonzero. However, as shown in column 3, if we ignore the $\alpha^2$ terms, we recover the strict probabilities.}
\label{Proba}
\end{table*}

We begin by considering donor and acceptor sites that are randomly assigned on a square lattice on each surface and defined by three lengthscales. The first one is the side length of the whole surface $L$. The second one is the size of a microscopic site $l_0$, which can donate/accept one elementary charge $e$. Finally, a third, intermediate lengthscale must be included to obtain realistic values for charge density~\cite{grosjean_quantitatively_2020}. It is the correlation length between sites $l$, which corresponds to the typical size of features on the mosaic. Once a donor site touches an acceptor site, there is a probability $\alpha$ that the transfer occurs. We will routinely express $L$ and $l$ in units of $l_0$, and charge in units of $e$. Before any contacts, the donor probability on a given grid M is $p_{\mathrm{M}}^0$, and the corresponding acceptor probability is $s_{\mathrm{M}}^0=1-p_{\mathrm{M}}^0$.

Let us first consider a single contact between two surfaces, meaning that $p_{\mathrm{M}} = p_{\mathrm{M}}^0$. Denoting the surfaces $A$ and $B$, we write the donor probabilities $p_{\mathrm{A}}$ and $p_{\mathrm{B}}$, and acceptor probabilities $s_{\mathrm{A}} = 1-p_{\mathrm{A}}$ and $s_{\mathrm{B}}=1-p_{\mathrm{B}}$. We first look at the case where $l=l_0$. Focusing on an individual pair of opposing sites, three situations can occur during contact. First, there can be transfer from A to B, which occurs with probability $p_{\mathrm{A}} (1-p_{\mathrm{B}})\,\alpha$. Second, there can be transfer from B to A, which occurs with probability $p_{\mathrm{B}} (1-p_{\mathrm{A}})\,\alpha$. Finally, there can be no transfer at all, which has probability $1 - (p_{\mathrm{A}} +p_{\mathrm{B}} - 2p_{\mathrm{A}}p_{\mathrm{B}})\,\alpha$. These values are summarized in the first column of Table~\ref{Proba}. For large surfaces, \textit{i.e.}~when $L\gg l_0$, the probability of the net charge transfer from A$\rightarrow$B is Gaussian with mean $e L^2/l_0^2\,p_{\mathrm{A}} \left(1-p_{\mathrm{B}} \right)\alpha$ and width $e L/l_0 \sqrt{p_{\mathrm{A}} (1-p_{\mathrm{B}})\,\alpha\,(1-p_{\mathrm{A}} (1-p_{\mathrm{B}})\,\alpha)}$. The expression for A$\leftarrow$B transfer is the same with $p_{\mathrm{A}}$ and $p_{\mathrm{B}}$ permuted. 

Determining the distribution for the \textit{total} charge exchange is more complicated as it involves the difference of A$\rightarrow$B and A$\leftarrow$B, and at a particular site transfer both ways cannot technically happen simultaneously. A helpful simplification comes if we approximate the A$\rightarrow$B and A$\leftarrow$B events as independent. The probabilities for this scenario are shown in column 2 of Table~\ref{Proba}, where there is now a possibility of simultaneous transfer A$\rightleftarrows$B. Considering experiments suggest $\alpha \ll 1$~\cite{apodaca_contact_2010}, and knowing that $p_{\mathrm{A}} \leq 1$ and $p_{\mathrm{B}} \leq 1$, we make the assumption that $\alpha^2$ terms can be ignored. As shown in column 3 of Table~\ref{Proba}, the relaxed case with independent events is then approximately equal to the strict case of column 1.

Now treating the distributions of A$\rightarrow$B and A$\leftarrow$B transfer as independent Gaussians, the distribution for the total charge exchanged after one contact $Q_1$ is also a Gaussian, with mean 
\begin{equation}
  \overline{Q_1}_{\,l=l_0} = e \alpha \frac{L^2}{l_0^2} \left(p_{\mathrm{A}} - p_{\mathrm{B}}\right)
  \label{eq:mu_l0}
\end{equation}
and standard deviation
\begin{equation}
  \sigma_{l=l_0} \approx e \frac{L}{l_0} \sqrt{\alpha \left(p_{\mathrm{A}} + p_{\mathrm{B}} - 2p_{\mathrm{A}} p_{\mathrm{B}} \right)}.
  \label{eq:sigma_l0}
\end{equation}
Again, these expressions assume we can ignore $\alpha^2$ terms in $\sigma$ and take the convention that the positive direction for charge exchange is A$\rightarrow$B.

The case where $l>l_0$ is treated similarly, but considering a rescaled matrix containing large patches of $n = (l/l_0)^2$ sites. The distance $l$ can be understood as the typical size of a patch, the average distance at which sites become uncorrelated or, equivalently, the first zero of the spatial correlation function~\cite{grosjean_quantitatively_2020}. To calculate the distribution, we will not immediately consider charge transfer, and will first consider instead the number of patches of $n$ donors on A facing patches of $n$ acceptors on B, which we denote $N_{\rightarrow}$. There is probability $p_{\mathrm{A}}(1-p_{\mathrm{B}})$ that a donor patch on A would face an acceptor patch on B. Assuming $L\gg l$, the values of $N_{\rightarrow}$ follow a normal distribution with mean $\mu_{\rightarrow} = L^2/l^2 p_{\mathrm{A}} (1-p_{\mathrm{B}})$ and width $\sigma_{\rightarrow} = L/l \sqrt{p_{\mathrm{A}} (1-p_{\mathrm{B}}) (1-p_{\mathrm{A}} (1-p_{\mathrm{B}}))}$. Denoting $N_{\leftarrow}$ the number donor patches on B facing acceptor patches on A, we find the same expressions for $\mu_{\leftarrow}$ and $\sigma_{\leftarrow}$ with A and B permuted. Treating the two distributions as independent, the distribution for $N_{\rightarrow} - N_{\leftarrow}$ is also a Gaussian with mean $\mu_{\rightarrow}-\mu_{\leftarrow}$ and standard deviation $\sqrt{\sigma_{\rightarrow}^2+\sigma_{\leftarrow}^2}$.

We note that each patch can transfer or receive a maximum of $n$ elementary charges. If $n\gg 1$, then the actual number of charges transferred by a donor patch is close to $\alpha n$, with fluctuations $\sqrt{n \alpha (1-\alpha)} \ll \alpha n$. This means that $\alpha$ is treated slightly differently here, determining the average transfer rate between patches. We find the same expression for $\overline{Q_1}_{\,l>l_0}$ as the one from Eq.~(\ref{eq:mu_l0}), whereas the standard deviation can be written
\begin{equation}
\begin{split}
  \sigma_{l>l_0} \approx e \frac{L l}{l_0^2} \alpha \sqrt{\left(p_{\mathrm{A}} + p_{\mathrm{B}} - 2p_{\mathrm{A}} p_{\mathrm{B}} \right)}
\end{split}
\end{equation}
assuming that $p_{\mathrm{A}} \approx p_{\mathrm{B}}$. In this case, the differences with Eq.~(\ref{eq:sigma_l0}) lay in the exponent $1/2$ of $\alpha$, and a factor $l/l_0$.

Finally, we express the donor probabilities in terms of their difference $\delta = p_{\mathrm{B}} - p_{\mathrm{A}}$ and average $p=(p_{\mathrm{A}}+p_{\mathrm{B}})/2$. It follows that the charge exchanged after one contact follows a Gaussian distribution with mean
\begin{equation}
  \overline{Q_1} = -e \alpha \frac{L^2}{l_0^2} \delta \equiv \mu
  \label{eq:mu}
\end{equation}
and width
\begin{equation}
  \sigma \approx e \frac{L l}{l_0^2} \alpha^m \sqrt{2 \left( p\left(1-p\right)+\frac{\delta^2}{4} \right)}
  \label{eq:sigma}
\end{equation}
where $m=0.5$ for $l=l_0$ and $1$ for $l>l_0$. While $\delta$ appears in $\sigma$, its most noticeable effect compared to the known case of $p_{\mathrm{A}}=p_{\mathrm{B}}$ is that the distribution is no longer centered around zero. If $\delta=0$, charge transfer is only powered by the statistical differences between two surfaces, which lead to a nonzero charge exchange magnitude $\overline{|Q_1|}=\sigma\sqrt{2/\pi}$. In the general case however, charge exchange magnitude follows a folded normal distribution which can be written
\begin{equation}
  \overline{\left|Q_1\right|} = \sigma \sqrt{\frac{2}{\pi}} \exp{\frac{-\mu^2}{2 \sigma^2}} + \mu \erf{\frac{\mu}{\sqrt{2 \sigma^2}}}
  \label{eq:abs}
\end{equation}
where $\erf{(x)}$ denotes the Gauss error function. One can easily show that Eqs.~(\ref{eq:mu}) through (\ref{eq:abs}) recover the results from Ref.~\cite{grosjean_quantitatively_2020} when $\delta=0$.

\begin{figure}
  \centering
  \includegraphics[width=8.6cm]{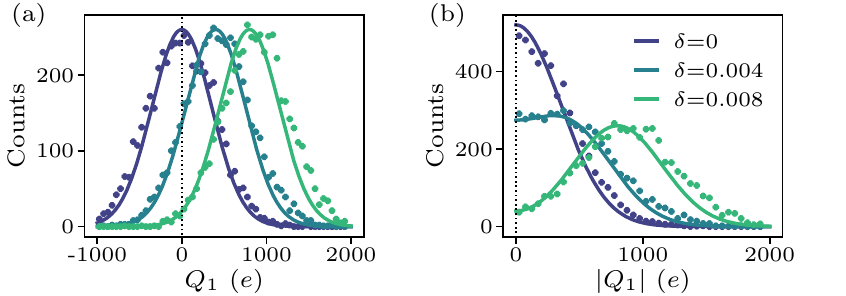}
  \includegraphics[width=8.6cm]{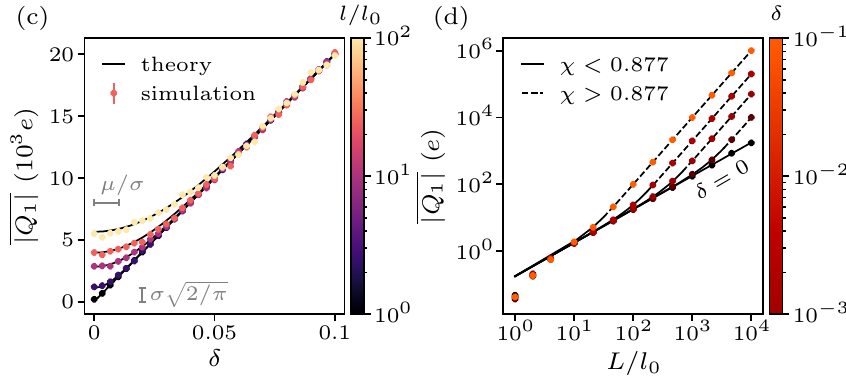}
  \caption{Model for donor probability asymmetry. (a) The charge exchanged during a single contact $Q_1$ follows a Gaussian distribution, no longer centered around $0$ when $\delta \neq 0$. Median and widths are given by Eqs.~(\ref{eq:mu}) and (\ref{eq:sigma}), respectively. Statistics are shown for 5000 pairs of surfaces with $p=0.5$, $\alpha=0.1$, $l=5\,l_0$ and $L=1000\,l_0$. (b) Charge exchange magnitude $|Q_1|$ follows a folded normal distribution. (c) Average magnitude $\overline{|Q_1|}$ is described by Eq.~(\ref{eq:abs}). The intercept $\sigma \sqrt{2/\pi}$ corresponds to the same-material case. The ratio $\mu/\sigma$, proportional to $l/l_0$, determines whether charging is powered by fluctuations or not. (d) Charging can be linear or quadratic in $L$, depending on whether charging is average-driven or variance-driven. The same system can go from linear to quadratic as $L$ increases. 
  }
  \label{fig2}
\end{figure}

To validate this analysis, we perform numerical simulations, where surfaces are generated and `contacts' performed using the method described in Ref.~\cite{grosjean_quantitatively_2020}. Features of size $l$ can be included by interpolating a coarse $L/l \times L/l$ matrix filled with random numbers between 0 and 1, then calculating the threshold value that leads to a binary field with the desired donor probability $p$~\cite{grosjean_quantitatively_2020}. Figure~\ref{fig2}(a) shows the distribution of charge transferred after contact between two square surfaces of side lengths $L=1000 \, l_0$, with different colors corresponding to different values of the donor density difference, $\delta$. Each distribution is determined from simulating 2000 pairs of such surfaces. As can be seen, introducing $\delta>0$ shifts the Gaussian profile, which is no longer centered around zero. The solid lines correspond to normal distributions whose means and widths are given by Eqs.~(\ref{eq:mu}) and (\ref{eq:sigma}). Figure~\ref{fig2}(b) shows the magnitude of charge exchange $|Q_1|$ for the same set of data. For small values of $\delta$, the distributions are folded, their average value $\overline{|Q_1|}$ being given by Eq.~(\ref{eq:abs}). As $\delta$ increases and the distributions move away from zero, they become essentially Gaussian.

The folded Gaussian and Gaussian distributions can be understood as two regimes, one where the standard deviation, $\sigma$, most significantly contributes to $\overline{|Q_1|}$, and the other where the average $\mu$, and therefore $\delta$, is primarily responsible. These variance-driven and average-driven regimes, at their extremes, can be understood as same-material CE with macroscopically identical samples, and CE between two different materials, respectively. One possible way to identify where the regime change occurs is to compare the two terms of Eq.~(\ref{eq:abs}). Posing $\chi = \mu/\sigma$, the terms are equal if
\begin{equation}
  \sqrt{\frac{2}{\pi}} \exp{\frac{-\chi}{2}} = \chi \erf{\frac{\chi}{\sqrt{2}}}
  \label{eq:chi}
\end{equation}
which can be solved numerically to find $\chi \approx \pm 0.8769$. Note that $\chi$ is proportional to $l/l_0$ and inversely proportional to $L/l_0$, meaning that systems where correlations lengths are small compared to their overall size tend to be more average-dominated. Figure~\ref{fig2}(c) compares Eq.~(\ref{eq:abs}) with simulations. Each point corresponds to 1000 first contacts between two $1000 \times 1000$ grids, with $\alpha=1$ and $p=0.5$, averaged over 30 pairs of grids. Five values of $l/l_0$ are shown, namely 1, 20, 50, 70 and 100. The intercepts in $\delta=0$ correspond to the known case of $\overline{|Q_1|}=\sigma \sqrt{2/\pi}$. For large $\delta$, the effect of correlation length $l/l_0$ disappears as charge exchange is purely driven by average differences between surfaces. 

We can point to experiments whose outcomes are consistent with the scaling laws obtained for the variance-dominated and average-dominated regimes. For instance, we can note that $\mu \sim L^2$ and $\sigma \sim L$. In experiments with polymers, Apodaca \emph{et al.} measured $\overline{|Q_1|}$ as a function of $L$~\cite{apodaca_contact_2010}. When contacting Polydimethylsiloxane (PDMS) with either Polyvinyl chloride (PVC), stainless steel or copper, they found that $\overline{|Q_1|} \sim L^\eta$ with values of $\eta$ between 1.6 and 1.9, slightly below 2. With PDMS-PDMS contacts, however, they observed that $\eta$ dropped to approximately 1. Equation~(\ref{eq:abs}) predicts such linear and quadratic behaviors of $\overline{|Q_1|}$ with $L$, with a continuous transition between them when the contributions of $\mu$ and $\sigma$ are comparable. The fact that $\mu \sim L^2$ and $\sigma \sim L$ also means that the same materials could behave differently at different sizes, \textit{i.e.}~charging could be variance-dominated for smaller samples, but average-dominated for larger ones. Figure~\ref{fig2}(d) illustrates this, comparing simulations with Eqs.~(\ref{eq:abs}) and (\ref{eq:chi}). 


\begin{figure}
  \centering
  \includegraphics[width=8.6cm]{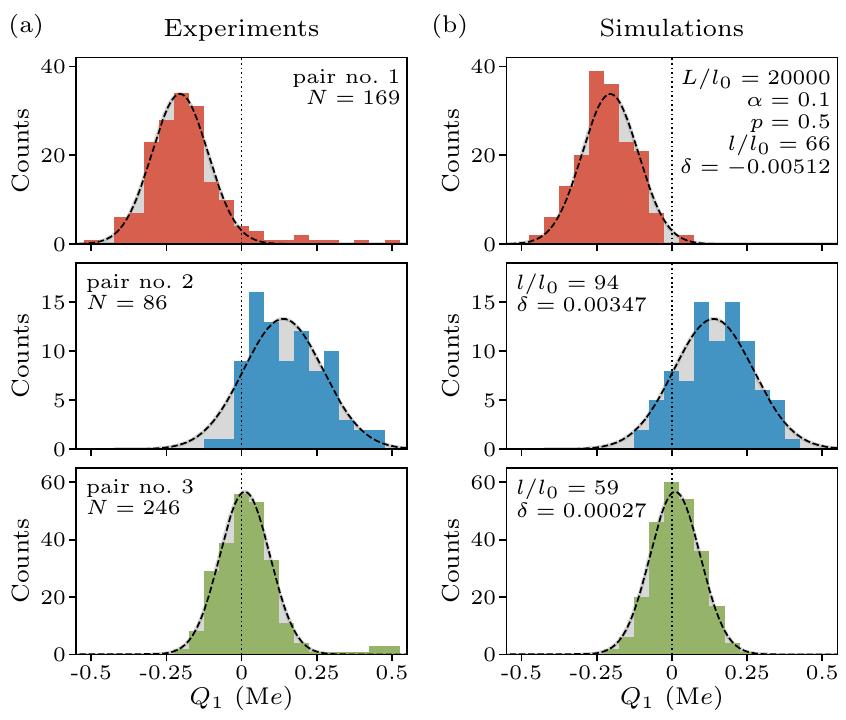}
  \caption{Comparison with experiments. (a) Single collision statistics measured experimentally, taken from Ref.~\cite{grosjean_single-collision_2023}, for three pairs of samples. The dashed line is a Gaussian fit. (b) Simulations based on donor/acceptor asymmetry proposed in this paper, performed with parameters inferred from the fits using Eqs.~(\ref{eq:mu}) and (\ref{eq:sigma}).  The model is able to explain the experimental data, with donor/acceptor donor density differences $\delta$ on the order of $< 1$\%. }
  \label{fig3}
\end{figure}

We can also use this framework to make sense of recent experiments~\cite{grosjean_single-collision_2023} where single-contact charge-transfer distributions were measured for same-material samples. Figure~\ref{fig3} shows three such distributions taken from Ref.~\cite{grosjean_single-collision_2023}, measured using acoustic levitation to perform collisions between a sphere and a plate, both made of fused silica. These distributions are not always centered around zero, meaning that there are global differences between samples which can potentially dominate over local fluctuations. Incorporating a difference in donor probability $\delta \neq 0$ in the model makes off-center distributions possible. We know from Ref.~\cite{grosjean_single-collision_2023} that $L \approx 20$~µm. Using Eqs.~(\ref{eq:mu}) and (\ref{eq:sigma}), we can extract $\mu$ and $\sigma$ from the distributions. We then translate those values in terms of $l/l_0$ and $\delta$ by making reasonable assumptions for the remaining parameters. Inspired by measurements from Refs.~\cite{apodaca_contact_2010} and \cite{baytekin_mosaic_2011}, we pose $\alpha=0.1$, $p=0.5$ and $l_0=1$~nm. From Eq.~(\ref{eq:sigma}), we find that
\begin{equation}
  l = \frac{\sigma l_0^2}{\alpha L \sqrt{2\left( p (1-p) + \frac{\delta^2}{4} \right)}},
\end{equation}
and from equ.~(\ref{eq:mu}), we get
\begin{equation}
  \delta = \frac{\mu l_0^2}{e \alpha L^2}.
\end{equation}
Assuming by convention that the distributions were measured for surface A and that the charge carriers are positive, we obtain the values for $l/l_0$ and $\delta$ shown in Fig.~\ref{fig3}(b). We can then use these parameters as inputs for numerical simulations which we contrast against the experiments, performing the same number of measurements $N$.

Thanks to the introduction of $\delta$, the simulations are able to reproduce the experimental results well, as illustrated by plotting the Gaussian fits from the experiment over the simulated data. The width of the distributions is relatively constant, leading to little variation of $l/l_0 \approx 73 \pm 15$, which is not too far from the size of the features observed in Ref.~\cite{baytekin_mosaic_2011}. We also note that $|\delta| = 0.003 \pm 0.002$, which is small compared to $p$. This means that very slight, but nonetheless global differences in donor/acceptor densities can significantly affect the outcome of experiments, in a way that is distinct from purely local fluctuations. The simulations from Fig.~\ref{fig3}(b) show that the generalized mosaic framework can produce realistic results, which could for example be incorporated into simulations of granular media. Nonetheless, we had to resort to assumptions for parameters that are difficult to access experimentally. Dedicated measurements could solve this issue. For instance, $\alpha$ can be obtained by repeating contacts if the location on the samples is fixed. Atomic force microscopy could be used to measure $l$ and perhaps estimate $p$, assuming such features can be observed on fused silica samples. The regime change from $|Q|\sim L$ to $|Q|\sim L^2$ from Fig.~\ref{fig2}(d), which depends on the ratio $\mu/\sigma$, could, if observed, be contrasted with independent measurements of $\mu$ and $\sigma$. Regime change or not, reproducing the $|Q|\sim L$ scaling from Ref.~\cite{apodaca_contact_2010} would be highly valuable. Considering that very small ($<$1\%) differences in donor/acceptor densities are sufficient to leave the variance-driven regime, it is possible that this regime would only exist in carefully controlled conditions with strictly identical samples. It is, however, difficult to estimate a plausible range for $\delta$ without knowing more about the nature of the donors/acceptors.

\section{Asymmetry during sliding}

\begin{figure}
  \centering
  \includegraphics[width=8.6cm]{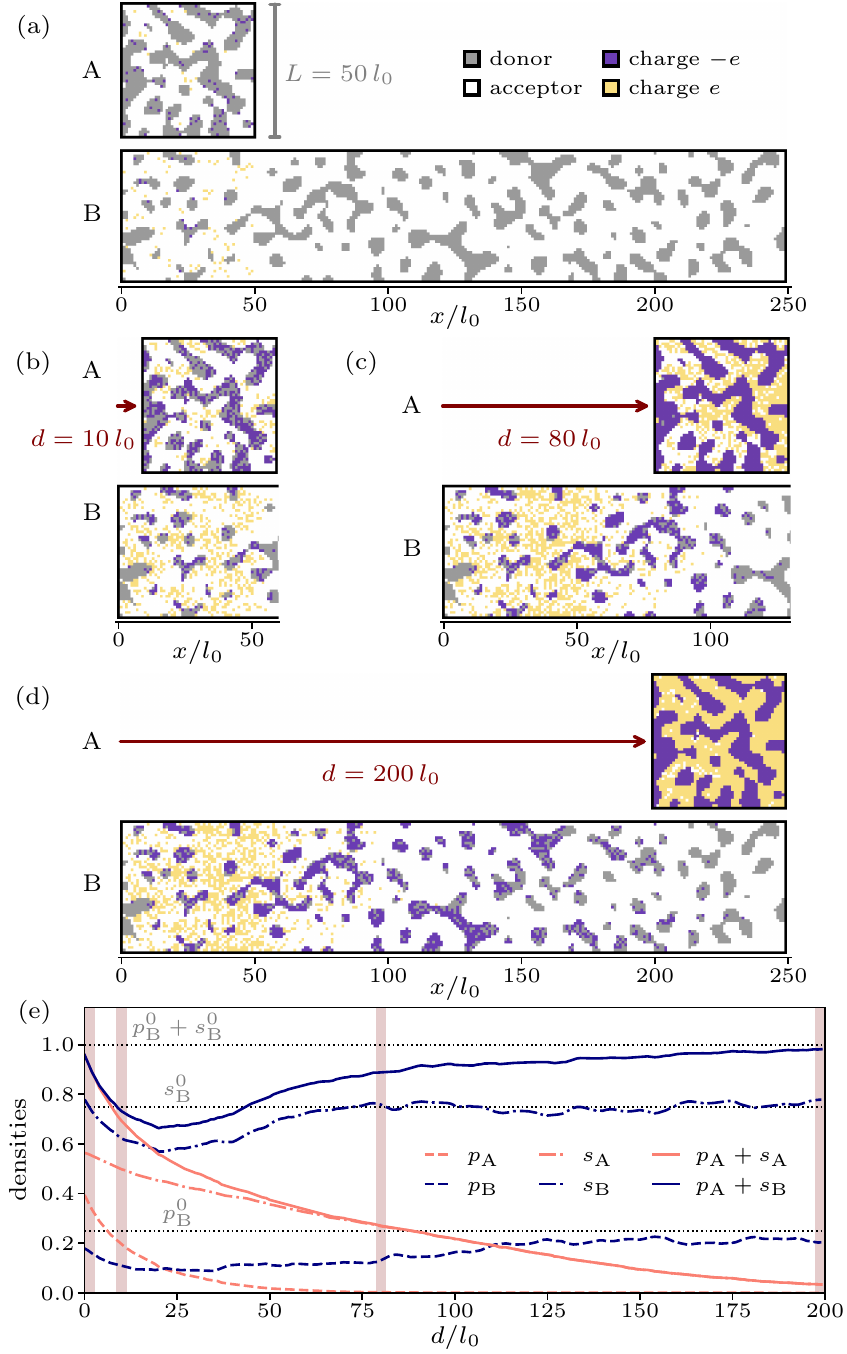}
  \caption{Asymmetry during sliding. (a) A square grid A with $50\times 50$ sites slides on a rectangular grid B with $50\times 250$ sites. Parameters are $\alpha=0.1$, $p_{\mathrm A}^0=0.45$, $p_{\mathrm B}^0=0.25$, and $l/l_0=4$. Every time sliding distance $d$ is increased by one site length $l_0$, a new contact occurs. After one contact, some donor and acceptor sites become charged and no longer contribute. (b) As grid A slides along B, it is exposed to new regions with unused sites. (c) Around $d=80\,l_0$, most donors on A have been used. Only the acceptor sites contribute, causing charging direction to reverse. (d) At $d=200\,l_0$, most acceptors on A have been used, and charging stops. (e) Densities $p_{\mathrm{A}}$ and $s_{\mathrm{A}}$ both decrease monotonously with $d$, but at different rates. On the other hand, the section of B facing A is continually renewed. At $d=200\,l_0$, we see that $p_{\mathrm{A}} \approx 0$ and $s_{\mathrm{A}} \approx 0$, but $s_{\mathrm{B}} \approx s_{\mathrm{B}}^0$ and $p_{\mathrm{B}} \approx p_{\mathrm{B}}^0$. Vertical lines indicate the values of $d$ which correspond to panels (a) through (d).
  }
  \label{fig4}
\end{figure}

We now show how thinking about donor/acceptor density differences can be used to treat another issue in CE, namely when a smaller area on one surface is slid over a larger area on another. We consider two samples of different size, divided into microscopic square sites of side length $l_0$. Sample A of size $w \times L$ is slid along a larger sample B of size $W \times L$, with $W > w$. We denote $d$ the distance over which A has slid on B. If we consider that A and B are fully in contact at the start and end of the sliding, we have $d \leq W - w$. The first contact occurs at sliding distance $d=0$. Subsequently, every time $d$ increases by $l_0$, a new contact occurs. Figure~\ref{fig4} illustrates sliding using a simulation where $w = L = 50\,l_0$ and $W = 250\,l_0$. 

During sliding, certain donors and acceptors will contribute and become inactive, leading donor and acceptor densities $p_{\mathrm A}$ and $s_{\mathrm A}$ to decrease with $d$. On B, however, the sliding continually brings new, unused sites. Note that we only consider the section of B which faces A at any given time when calculating $p_{\mathrm B}$ and $s_{\mathrm B}$. This means that, even in a situation where the donor/acceptor densities on A and B are initially equal, they will eventually become unequal---hence recreating the donor/acceptor asymmetry discussed in the previous section. In general, we have $p_{\mathrm{M}}(d, x)$ and $s_{\mathrm{M}}(d, x)$, where $x$ is the horizontal position on M along the sliding direction. We denote $p_{\mathrm M}^0$ and $s_{\mathrm M}^0 = 1-p_{\mathrm M}^0$ the initial donor and acceptor densities. Note that for $d>0$, we can no longer assume that $s_{\mathrm{M}}(d, x) = 1 - p_{\mathrm{M}}(d, x)$. In Fig.~\ref{fig4}(a), $d=0$ and only one contact has occurred. Most sites are still in their unused state (shown in white for donors and gray for acceptors).

Figure~\ref{fig4}(e) shows the evolution with $d$ of donor/acceptor densities on A and B. First, one can see that $p_{\mathrm{A}}$ and $s_{\mathrm{A}}$ decrease monotonously, but at different rates. This is also visible in Fig.~\ref{fig4}(c), where all donor sites on A have already transferred their charge (in dark purple), while there are still many acceptor sites available (in white). This can cause a reversal of the charging direction. Sample A is not uniform at this stage, with most of the remaining acceptor sites (in white) located toward the trailing edge, i.e. toward smaller $x$. The evolution of donor/acceptor densities on B is qualitatively different from A, as can be seen in Fig.~\ref{fig4}(e). After an initial decrease of $p_{\mathrm{B}}$ and $s_{\mathrm{B}}$, they slowly come back to approximately their initial values $p_{\mathrm{B}}^0$ and $s_{\mathrm{B}}^0$. This is because the part of B contributing to charge exchange is constantly renewed. Looking at B on Fig.~\ref{fig4}(d), we can see that the acceptors which have acquired charge (light yellow) are concentrated toward the beginning of the track, below $x\approx50\,l_0$, while the used up acceptors (dark purple) are spread over a much larger portion of the track, up to $x\approx200\,l_0$.

In normal, forward facing contacts, the number of sites contributing to charge exchange at every contact is the number of donors facing acceptors (and vice-versa) multiplied by the transfer probability $\alpha$. Writing the number of contacts $n$ as a continuous variable, one can show than the number of charges being exchanged at every step decreases like $\exp\left(-\alpha n\right)$~\cite{apodaca_contact_2010,grosjean_quantitatively_2020}. However, when sliding, the number of contacts $n$ experienced by a given site is now a function of the location on A or B, as discussed above. Every time A moves by $l/l_0$, the first column of sites in the direction of the sliding faces `new' sites on B. As a consequence, sites on the leading edge of A will initially experience more charge exchange than sites on the trailing edge.

The whole surface of A contributes to every contact, as every site on A experiences $n_{\mathrm{A}}=d/l_0$ contacts. On B, however, the number of contacts is a function of the position $x$ along the sliding direction, which can be written as the convolution product $n_{\mathrm{B}} \left(x\right) = \int_{0}^{x} H(w-\tau) H(d-(x-\tau)) \,\mathrm{d}\tau$ where $H$ is the Heaviside step function. This causes, in turn, $p_{\mathrm{B}}$ and $s_{\mathrm{B}}$ to depend on $x$, as well as $p_{\mathrm{A}}$ and $s_{\mathrm{A}}$ through the contacts with B, which makes the analytical calculation of $Q$ significantly more difficult in the general case.

The case $w=l_0$, however, is comparatively straightforward. Sample A becomes as thin as possible and the dependency on $x$ disappears. In this case, $p_{\mathrm{B}}$ and $s_{\mathrm{B}}$ remain constant, and every additional contact causes $p_{\mathrm{A}}$ to decrease by a factor $(1-s_{\mathrm{B}}^0)\,\alpha$, and $s_{\mathrm{A}}$ by a factor $(1-p_{\mathrm{B}}^0)\,\alpha$. To illustrate this, we write the evolution of $p_{\mathrm{A}}$ for $d={0, 1, 2 \dots n}$, keeping in mind that $s_{\mathrm{B}}=s_{\mathrm{B}}^0$ as the surface is continually renewed. At every contact, we subtract the number of transfers going from A to B. This leads to
\begin{equation*}
\begin{split}
p_{\mathrm{A}}(0) &= p_{\mathrm{A}}^0\\
p_{\mathrm{A}}(1) &= p_{\mathrm{A}}(0) - p_{\mathrm{A}}(0)\, s_{\mathrm{B}}^0\, \alpha = p_{\mathrm{A}}^0 \left( 1-s_{\mathrm{B}}^0\, \alpha \right)\\
p_{\mathrm{A}}(2) &= p_{\mathrm{A}}(1) - p_{\mathrm{A}}(1)\, s_{\mathrm{B}}^0\, \alpha = p_{\mathrm{A}}^0 \left( 1-2\,s_{\mathrm{B}}^0\, \alpha + \left(s_{\mathrm{B}}^0\, \alpha\right)^2 \right)\\
&~~\vdots\\
p_{\mathrm{A}}(n) &= p_{\mathrm{A}}^0 \left( 1-n\, s_{\mathrm{B}}^0\, \alpha \right) + \mathcal{O}(\alpha^2).
\end{split}
\end{equation*}
The reasoning is the same for $s_{\mathrm{A}}$. We ignore the $\alpha^2$ terms and consider $d$ as a continuous variable. When $d$ increases by a small increment $\delta d$, $p_{\mathrm{A}}$ increases by
\begin{equation*}
    \delta p_{\mathrm{A}} = - p_{\mathrm{A}} \left( 1 - s_{\mathrm{B}}^0 \right) \alpha \, \delta d
\end{equation*}
which we integrate to obtain
\begin{equation}
  p_{\mathrm{A}} \left( d \right) = p_{\mathrm{A}}^0 \exp{\left( - \alpha \left( 1 - p_{\mathrm{B}}^0 \right) \frac{d}{l_0} \right)} = p_{\mathrm{A}}^0 \exp{-\frac{d}{\lambda_p}}
\label{eq:pA}.
\end{equation}
Similarly, we find
\begin{equation}
  s_{\mathrm{A}} \left( d \right) = \left( 1 - p_{\mathrm{A}}^0 \right) \exp{\left( - \alpha \, p_{\mathrm{B}}^0 \frac{d}{l_0} \right)} = s_{\mathrm{A}}^0 \exp{-\frac{d}{\lambda_s}}
\label{eq:sA}
\end{equation}
where $\lambda_p = l_0 / \alpha \,(1-p_{\mathrm{B}}^0)$ and $\lambda_s = l_0 / \alpha \,p_{\mathrm{B}}^0$ are the characteristic sliding lengths over which donors and acceptors on A get depleted. 

\begin{figure}
  \centering
  \includegraphics[width=8.6cm]{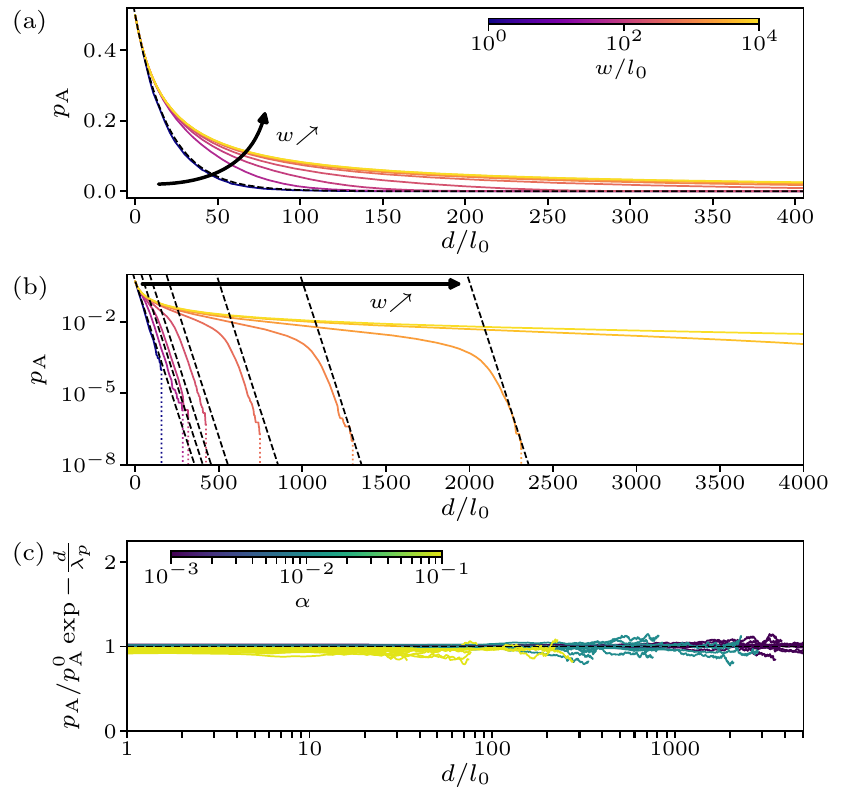}
  \caption{Simple sliding model. (a) The decrease of $p_{\mathrm{A}}$ follows Eq.~(\ref{eq:pA}) (dashed line) only if grid A is infinitely narrow, \textit{i.e.}~when $w=l_0$. We start with a $1 \times 10000$ grid and increase the width progressively up to $10000 \times 10000$. (b) On a semi-log scale, one can see that Eq.~(\ref{eq:pA}) works for sliding distances $d \ll L$, as A is still approximately uniform, and for $d \geq w$. (c) We verify Eq.~(\ref{eq:pA}) for multiple values of $\alpha$, $p$ and $\delta$, but keeping $L=10000\,l_0$, $w=l_0$ and $l=l_0$.
  }
  \label{fig5}
\end{figure}

Figure~\ref{fig5} compares Eq.~(\ref{eq:pA}) with numerical simulations, with $L=W=10000\,l_0$, $p_{\mathrm{A}}^0 = s_{\mathrm{A}}^0 = p_{\mathrm{B}}^0 = p_{\mathrm{B}}^0 = 0.5$ and $l=l_0$. The width $w$ is increased from $l_0$ to $L$. One can see that when $w>l_0$ and $d>l_0$, we depart from the exponential behavior. The decrease of $p_{\mathrm{A}}$ is slowed down by the presence of used up sites on B. In other words, we locally have $p_{\mathrm{B}} < p_{\mathrm{B}}^0$ and $s_{\mathrm{B}} < s_{\mathrm{B}}^0$, leading $\lambda_p$ and $\lambda_s$ to effectively increase. As can be seen on Fig.~\ref{fig5}(b), this causes the exponential to plateau until about $d \geq w$, where the remaining active sites on A become sufficiently sparse so that we once more have $p_{\mathrm{B}} \approx p_{\mathrm{B}}^0$ and $s_{\mathrm{B}} \approx s_{\mathrm{B}}^0$. Figure~\ref{fig5}(c) compares Eq.~(\ref{eq:pA}) for a wider range of parameters. We see a good agreement for any combination of $\delta$ and $p$, granted that $\alpha\ll 1$ and $l\ll L$. The curves are plotted until $p_{\mathrm{A}}$ has decreased by 99\%, as fluctuations become large when only a few sites remain.

From Eqs.~(\ref{eq:pA}) and (\ref{eq:sA}), we see that if $p_{\mathrm{B}} = 0.5$, then donors and acceptors on A are depleted at exactly the same rate. If $p_{\mathrm{B}} \neq 0.5$, there can exist a sliding distance $d^*$ such that $p_{\mathrm{A}} \left( d^* \right) = s_{\mathrm{A}} \left( d^* \right)$. From Eqs.~(\ref{eq:pA}) and (\ref{eq:sA}), we get
\begin{equation}
   d^* = \frac{\alpha \, l_0}{2 p_{\mathrm{B}} - 1} \ln{\left(\frac{1-p_{\mathrm{A}}^0}{p_{\mathrm{A}}^0}\right)}.
\label{eq:dcross}
\end{equation}
This means that past a certain sliding distance, the average direction of charge exchange can reverse. Note that $d^*$ is positive only if either $p_{\mathrm{A}} > 0.5$ and $p_{\mathrm{B}} < 0.5$, or $p_{\mathrm{A}} < 0.5$ and $p_{\mathrm{B}} > 0.5$. Such reversals were observed experimentally when sliding various materials on glass~\cite{shaw_tribo-electricity_1928} or with same-material samples~\cite{lowell_triboelectrification_1986}, but were generally attributed to the presence of contamination on the samples' surfaces. However, Eq.~(\ref{eq:dcross}) shows that reversals can naturally arise as a consequence of different donor/acceptor probabilities on the two samples. 

\begin{figure}
  \centering
  \includegraphics[width=8.6cm]{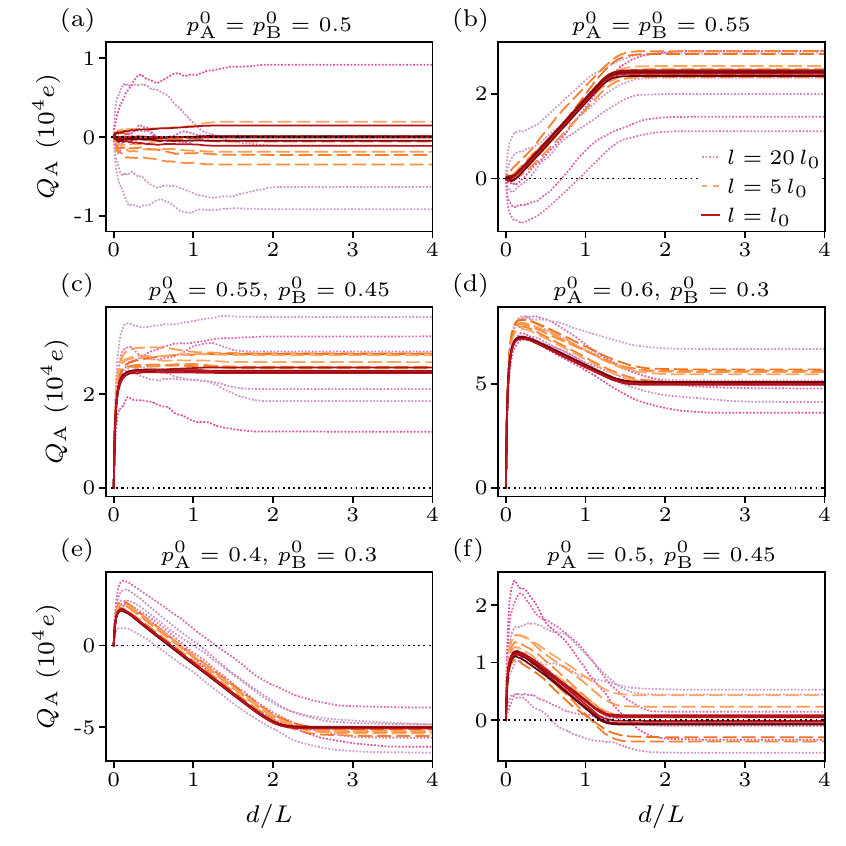}
  \caption{Reversals of charging direction and sign during sliding. We measure the charge of the square grid A as a function of sliding distance $d/l_0$, with $\alpha=0.1$, $L=500\,l_0$, and either $l=5\,l_0$ (solid red lines), $l=5\,l_0$ (dashed orange lines) or $l=20\,l_0$ (dotted purple lines). We show six trajectories for each set of parameters to illustrate inherent variability. (a,~b) When $p_{\mathrm{A}}^0 = p_{\mathrm{B}}^0 = p$, we observe that (a) $Q_{\mathrm{A}}=0$ on average when $p=0.5$, (b) $Q_{\mathrm{A}}$ increases (resp. decreases) monotonously for $p>0.5$ (resp. $p<0.5$). (c-f) When $p_{\mathrm{A}}^0 \neq p_{\mathrm{B}}^0$, a much wider range of charge behaviors can occur. This includes (c) monotonous charging, (d) non-monotonous charging with no zero-crossing, (e) zero-crossings or (f) combinations of those depending on random fluctuations, which increase with $l/l_0$.
  }
  \label{fig6}
\end{figure}

By running simulations, we can show that the necessary condition that $p_{\mathrm{A}}^0 \lessgtr 0.5$ and $p_{\mathrm{B}}^0 \gtrless 0.5$ can be somewhat relaxed for wider samples $w>l_0$. Indeed, as $p_{\mathrm{A}}$ and $p_{\mathrm{B}}$ become functions of the position on A and B, the condition for the reversal of charging direction can be met locally. In Fig.~\ref{fig6}, we measure the charge of A during sliding for various values of $p_{\mathrm{A}}^0$ and $p_{\mathrm{B}}^0$. The different behaviors observed in Refs.~\cite{shaw_tribo-electricity_1928,lowell_triboelectrification_1986} are reproduced, including reversal of charging direction (non-monotonous charging) and reversal of charge (zero-crossings). In general, non-monotonous charging only occurs if $p_{\mathrm{A}}^0 \neq p_{\mathrm{B}}^0$. For larger values of lengthscale $l/l_0$, the larger fluctuations between surfaces can also cause that condition to be met.

\section{Discussion and conclusions}

Looking at the literature, it is still unclear whether the patterns of positive and negative voltage sometimes observed after contact are a sign of actual charge donor/acceptor mosaics~\cite{baytekin_mosaic_2011}, a consequence of discharge during separation~\cite{sobolev_charge_2022}, or something else entirely. Regardless, mosaic models naturally follow from the assumption that charging is not a perfectly uniform process, and provide a useful conceptual tool for analytical and numerical studies. By accounting for global differences between samples, we extend this framework from strictly identical to different-material samples, and anything in between. Our model shows that whether charging is driven by global differences or local fluctuations depends not only on material properties, but on scale. Local fluctuations tend to matter more for smaller samples, and the transition might be within reach of experimental observations. 

Accounting for differences in donor/acceptor densities allows us to not only reproduce same-material CE measurements~\cite{grosjean_single-collision_2023}, but also make sense of contact sliding experiments~\cite{shaw_tribo-electricity_1928,lowell_triboelectrification_1986}. Combined with the asymmetric geometry due to the sliding, donor/acceptor asymmetries are sufficient to explain the sign flips that were observed experimentally, as an alternative explanation to impurities or contamination. 

One might argue that the flexibility of mosaic models is also their weakness. Some parameters are difficult to access experimentally, which so far has prevented the field from drawing conclusions about the underlying mechanism. However, we suggest experiments based on our results which might allow to build a fuller picture of CE. Repeatedly contacting two samples, resetting between contacts by discharging, provides the mean $\mu$ and standard deviation $\sigma$ from Eqs.~(\ref{eq:mu}) and (\ref{eq:sigma}), as established here with the data from Ref.~\cite{grosjean_single-collision_2023}. If multiple contacts are performed on the same location, one can determine transfer probability $\alpha$. Varying contact area over a wide enough range would allow to verify Eq.~(\ref{eq:abs}) by providing an independent measurement of the ratio $\mu/\sigma$. Deviations from the law could reveal information about $l$ and $l_0$, or at the very least provide an upper bound. Finally, sliding experiments provide an underutilized tool to test CE theories, as the charging profiles depend on the donor/acceptor probabilities of the two surfaces in non-trivial ways. Performing same-material and different-material sliding experiments in a variety of conditions, including changing the aspect ratio of the contact area, and noting if and where the sign flips occur could be a new way to estimate donor/acceptor probabilities to probe the underlying principles governing tribocharging. 

\section*{Acknowledgements}

This project has received funding from the European Research Council Grant Agreement No.~949120 and from the European Union's Horizon 2020 research and innovation program under the Marie Sk\l{}odowska-Curie Grant Agreement No.~754411.

\bibliography{references}

\begin{thebibliography}{31}%
\makeatletter
\providecommand \@ifxundefined [1]{%
 \@ifx{#1\undefined}
}%
\providecommand \@ifnum [1]{%
 \ifnum #1\expandafter \@firstoftwo
 \else \expandafter \@secondoftwo
 \fi
}%
\providecommand \@ifx [1]{%
 \ifx #1\expandafter \@firstoftwo
 \else \expandafter \@secondoftwo
 \fi
}%
\providecommand \natexlab [1]{#1}%
\providecommand \enquote  [1]{``#1''}%
\providecommand \bibnamefont  [1]{#1}%
\providecommand \bibfnamefont [1]{#1}%
\providecommand \citenamefont [1]{#1}%
\providecommand \href@noop [0]{\@secondoftwo}%
\providecommand \href [0]{\begingroup \@sanitize@url \@href}%
\providecommand \@href[1]{\@@startlink{#1}\@@href}%
\providecommand \@@href[1]{\endgroup#1\@@endlink}%
\providecommand \@sanitize@url [0]{\catcode `\\12\catcode `\$12\catcode
  `\&12\catcode `\#12\catcode `\^12\catcode `\_12\catcode `\%12\relax}%
\providecommand \@@startlink[1]{}%
\providecommand \@@endlink[0]{}%
\providecommand \url  [0]{\begingroup\@sanitize@url \@url }%
\providecommand \@url [1]{\endgroup\@href {#1}{\urlprefix }}%
\providecommand \urlprefix  [0]{URL }%
\providecommand \Eprint [0]{\href }%
\providecommand \doibase [0]{https://doi.org/}%
\providecommand \selectlanguage [0]{\@gobble}%
\providecommand \bibinfo  [0]{\@secondoftwo}%
\providecommand \bibfield  [0]{\@secondoftwo}%
\providecommand \translation [1]{[#1]}%
\providecommand \BibitemOpen [0]{}%
\providecommand \bibitemStop [0]{}%
\providecommand \bibitemNoStop [0]{.\EOS\space}%
\providecommand \EOS [0]{\spacefactor3000\relax}%
\providecommand \BibitemShut  [1]{\csname bibitem#1\endcsname}%
\let\auto@bib@innerbib\@empty
\bibitem [{\citenamefont {Rescaglio}\ \emph {et~al.}(2017)\citenamefont
  {Rescaglio}, \citenamefont {Schockmel}, \citenamefont {Vandewalle},\ and\
  \citenamefont {Lumay}}]{rescaglio_combined_2017}%
  \BibitemOpen
  \bibfield  {author} {\bibinfo {author} {\bibfnamefont {A.}~\bibnamefont
  {Rescaglio}}, \bibinfo {author} {\bibfnamefont {J.}~\bibnamefont
  {Schockmel}}, \bibinfo {author} {\bibfnamefont {N.}~\bibnamefont
  {Vandewalle}},\ and\ \bibinfo {author} {\bibfnamefont {G.}~\bibnamefont
  {Lumay}},\ }\bibfield  {title} {\bibinfo {title} {Combined effect of moisture
  and electrostatic charges on powder flow},\ }\href
  {https://doi.org/10.1051/epjconf/201714013009} {\bibfield  {journal}
  {\bibinfo  {journal} {EPJ Web Conf.}\ }\textbf {\bibinfo {volume} {140}},\
  \bibinfo {pages} {13009} (\bibinfo {year} {2017})}\BibitemShut {NoStop}%
\bibitem [{\citenamefont {Cimarelli}\ \emph {et~al.}(2022)\citenamefont
  {Cimarelli}, \citenamefont {Behnke}, \citenamefont {Genareau}, \citenamefont
  {Harper},\ and\ \citenamefont {Van~Eaton}}]{cimarelli_volcanic_2022}%
  \BibitemOpen
  \bibfield  {author} {\bibinfo {author} {\bibfnamefont {C.}~\bibnamefont
  {Cimarelli}}, \bibinfo {author} {\bibfnamefont {S.}~\bibnamefont {Behnke}},
  \bibinfo {author} {\bibfnamefont {K.}~\bibnamefont {Genareau}}, \bibinfo
  {author} {\bibfnamefont {J.~M.}\ \bibnamefont {Harper}},\ and\ \bibinfo
  {author} {\bibfnamefont {A.~R.}\ \bibnamefont {Van~Eaton}},\ }\bibfield
  {title} {\bibinfo {title} {Volcanic electrification: recent advances and
  future perspectives},\ }\href {https://doi.org/10.1007/s00445-022-01591-3}
  {\bibfield  {journal} {\bibinfo  {journal} {Bull Volcanol}\ }\textbf
  {\bibinfo {volume} {84}},\ \bibinfo {pages} {78} (\bibinfo {year}
  {2022})}\BibitemShut {NoStop}%
\bibitem [{\citenamefont {Zhang}\ and\ \citenamefont
  {Zhou}(2020)}]{zhang_reconstructing_2020}%
  \BibitemOpen
  \bibfield  {author} {\bibinfo {author} {\bibfnamefont {H.}~\bibnamefont
  {Zhang}}\ and\ \bibinfo {author} {\bibfnamefont {Y.-H.}\ \bibnamefont
  {Zhou}},\ }\bibfield  {title} {\bibinfo {title} {Reconstructing the
  electrical structure of dust storms from locally observed electric field
  data},\ }\href {https://doi.org/10.1038/s41467-020-18759-0} {\bibfield
  {journal} {\bibinfo  {journal} {Nat Commun}\ }\textbf {\bibinfo {volume}
  {11}},\ \bibinfo {pages} {5072} (\bibinfo {year} {2020})}\BibitemShut
  {NoStop}%
\bibitem [{\citenamefont {Steinpilz}\ \emph {et~al.}(2020)\citenamefont
  {Steinpilz}, \citenamefont {Joeris}, \citenamefont {Jungmann}, \citenamefont
  {Wolf}, \citenamefont {Brendel}, \citenamefont {Teiser}, \citenamefont
  {Shinbrot},\ and\ \citenamefont {Wurm}}]{steinpilz_electrical_2020}%
  \BibitemOpen
  \bibfield  {author} {\bibinfo {author} {\bibfnamefont {T.}~\bibnamefont
  {Steinpilz}}, \bibinfo {author} {\bibfnamefont {K.}~\bibnamefont {Joeris}},
  \bibinfo {author} {\bibfnamefont {F.}~\bibnamefont {Jungmann}}, \bibinfo
  {author} {\bibfnamefont {D.}~\bibnamefont {Wolf}}, \bibinfo {author}
  {\bibfnamefont {L.}~\bibnamefont {Brendel}}, \bibinfo {author} {\bibfnamefont
  {J.}~\bibnamefont {Teiser}}, \bibinfo {author} {\bibfnamefont
  {T.}~\bibnamefont {Shinbrot}},\ and\ \bibinfo {author} {\bibfnamefont
  {G.}~\bibnamefont {Wurm}},\ }\bibfield  {title} {\bibinfo {title} {Electrical
  charging overcomes the bouncing barrier in planet formation},\ }\href
  {https://doi.org/10.1038/s41567-019-0728-9} {\bibfield  {journal} {\bibinfo
  {journal} {Nat. Phys.}\ }\textbf {\bibinfo {volume} {16}},\ \bibinfo {pages}
  {225} (\bibinfo {year} {2020})}\BibitemShut {NoStop}%
\bibitem [{\citenamefont {Lacks}\ and\ \citenamefont
  {Shinbrot}(2019)}]{lacks_long-standing_2019}%
  \BibitemOpen
  \bibfield  {author} {\bibinfo {author} {\bibfnamefont {D.~J.}\ \bibnamefont
  {Lacks}}\ and\ \bibinfo {author} {\bibfnamefont {T.}~\bibnamefont
  {Shinbrot}},\ }\bibfield  {title} {\bibinfo {title} {Long-standing and
  unresolved issues in triboelectric charging},\ }\href
  {https://doi.org/10.1038/s41570-019-0115-1} {\bibfield  {journal} {\bibinfo
  {journal} {Nat. Rev. Chem.}\ }\textbf {\bibinfo {volume} {3}},\ \bibinfo
  {pages} {465} (\bibinfo {year} {2019})}\BibitemShut {NoStop}%
\bibitem [{\citenamefont {Shaw}(1926)}]{shaw_electrical_1926}%
  \BibitemOpen
  \bibfield  {author} {\bibinfo {author} {\bibfnamefont {P.~E.}\ \bibnamefont
  {Shaw}},\ }\bibfield  {title} {\bibinfo {title} {The {Electrical} {Charges}
  from {Like} {Solids}},\ }\href {https://doi.org/10.1038/118659c0} {\bibfield
  {journal} {\bibinfo  {journal} {Nature}\ }\textbf {\bibinfo {volume} {118}},\
  \bibinfo {pages} {659} (\bibinfo {year} {1926})}\BibitemShut {NoStop}%
\bibitem [{\citenamefont {Lowell}(1975)}]{lowell_contact_1975}%
  \BibitemOpen
  \bibfield  {author} {\bibinfo {author} {\bibfnamefont {J.}~\bibnamefont
  {Lowell}},\ }\bibfield  {title} {\bibinfo {title} {Contact electrification of
  metals},\ }\href {https://doi.org/10.1088/0022-3727/8/1/013} {\bibfield
  {journal} {\bibinfo  {journal} {J. Phys. D: Appl. Phys.}\ }\textbf {\bibinfo
  {volume} {8}},\ \bibinfo {pages} {53} (\bibinfo {year} {1975})}\BibitemShut
  {NoStop}%
\bibitem [{\citenamefont {Zhang}\ \emph {et~al.}(2019)\citenamefont {Zhang},
  \citenamefont {Chen}, \citenamefont {Jiang}, \citenamefont {Lim},\ and\
  \citenamefont {Soh}}]{zhang_rationalizing_2019}%
  \BibitemOpen
  \bibfield  {author} {\bibinfo {author} {\bibfnamefont {X.}~\bibnamefont
  {Zhang}}, \bibinfo {author} {\bibfnamefont {L.}~\bibnamefont {Chen}},
  \bibinfo {author} {\bibfnamefont {Y.}~\bibnamefont {Jiang}}, \bibinfo
  {author} {\bibfnamefont {W.}~\bibnamefont {Lim}},\ and\ \bibinfo {author}
  {\bibfnamefont {S.}~\bibnamefont {Soh}},\ }\bibfield  {title} {\bibinfo
  {title} {Rationalizing the {Triboelectric} {Series} of {Polymers}},\ }\href
  {https://doi.org/10.1021/acs.chemmater.8b04526} {\bibfield  {journal}
  {\bibinfo  {journal} {Chem. Mater.}\ }\textbf {\bibinfo {volume} {31}},\
  \bibinfo {pages} {1473} (\bibinfo {year} {2019})}\BibitemShut {NoStop}%
\bibitem [{\citenamefont {Apodaca}\ \emph {et~al.}(2010)\citenamefont
  {Apodaca}, \citenamefont {Wesson}, \citenamefont {Bishop}, \citenamefont
  {Ratner},\ and\ \citenamefont {Grzybowski}}]{apodaca_contact_2010}%
  \BibitemOpen
  \bibfield  {author} {\bibinfo {author} {\bibfnamefont {M.}~\bibnamefont
  {Apodaca}}, \bibinfo {author} {\bibfnamefont {P.}~\bibnamefont {Wesson}},
  \bibinfo {author} {\bibfnamefont {K.}~\bibnamefont {Bishop}}, \bibinfo
  {author} {\bibfnamefont {M.}~\bibnamefont {Ratner}},\ and\ \bibinfo {author}
  {\bibfnamefont {B.}~\bibnamefont {Grzybowski}},\ }\bibfield  {title}
  {\bibinfo {title} {Contact {Electrification} between {Identical}
  {Materials}},\ }\href {https://doi.org/10.1002/anie.200905281} {\bibfield
  {journal} {\bibinfo  {journal} {Angew. Chem. Int. Ed.}\ }\textbf {\bibinfo
  {volume} {49}},\ \bibinfo {pages} {946} (\bibinfo {year} {2010})}\BibitemShut
  {NoStop}%
\bibitem [{\citenamefont {Knorr}(2011)}]{knorr_squeezing_2011}%
  \BibitemOpen
  \bibfield  {author} {\bibinfo {author} {\bibfnamefont {N.}~\bibnamefont
  {Knorr}},\ }\bibfield  {title} {\bibinfo {title} {Squeezing out hydrated
  protons: low-frictional-energy triboelectric insulator charging on a
  microscopic scale},\ }\href {https://doi.org/10.1063/1.3592522} {\bibfield
  {journal} {\bibinfo  {journal} {AIP Advances}\ }\textbf {\bibinfo {volume}
  {1}},\ \bibinfo {pages} {022119} (\bibinfo {year} {2011})}\BibitemShut
  {NoStop}%
\bibitem [{\citenamefont {Baytekin}\ \emph {et~al.}(2011)\citenamefont
  {Baytekin}, \citenamefont {Patashinski}, \citenamefont {Branicki},
  \citenamefont {Baytekin}, \citenamefont {Soh},\ and\ \citenamefont
  {Grzybowski}}]{baytekin_mosaic_2011}%
  \BibitemOpen
  \bibfield  {author} {\bibinfo {author} {\bibfnamefont {H.~T.}\ \bibnamefont
  {Baytekin}}, \bibinfo {author} {\bibfnamefont {A.~Z.}\ \bibnamefont
  {Patashinski}}, \bibinfo {author} {\bibfnamefont {M.}~\bibnamefont
  {Branicki}}, \bibinfo {author} {\bibfnamefont {B.}~\bibnamefont {Baytekin}},
  \bibinfo {author} {\bibfnamefont {S.}~\bibnamefont {Soh}},\ and\ \bibinfo
  {author} {\bibfnamefont {B.~A.}\ \bibnamefont {Grzybowski}},\ }\bibfield
  {title} {\bibinfo {title} {The {Mosaic} of {Surface} {Charge} in {Contact}
  {Electrification}},\ }\href {https://doi.org/10.1126/science.1201512}
  {\bibfield  {journal} {\bibinfo  {journal} {Science}\ }\textbf {\bibinfo
  {volume} {333}},\ \bibinfo {pages} {308} (\bibinfo {year}
  {2011})}\BibitemShut {NoStop}%
\bibitem [{\citenamefont {Sow}\ \emph {et~al.}(2012)\citenamefont {Sow},
  \citenamefont {Lacks},\ and\ \citenamefont
  {Mohan~Sankaran}}]{sow_dependence_2012}%
  \BibitemOpen
  \bibfield  {author} {\bibinfo {author} {\bibfnamefont {M.}~\bibnamefont
  {Sow}}, \bibinfo {author} {\bibfnamefont {D.~J.}\ \bibnamefont {Lacks}},\
  and\ \bibinfo {author} {\bibfnamefont {R.}~\bibnamefont {Mohan~Sankaran}},\
  }\bibfield  {title} {\bibinfo {title} {Dependence of contact electrification
  on the magnitude of strain in polymeric materials},\ }\href
  {https://doi.org/10.1063/1.4761967} {\bibfield  {journal} {\bibinfo
  {journal} {J. Appl. Phys.}\ }\textbf {\bibinfo {volume} {112}},\ \bibinfo
  {pages} {084909} (\bibinfo {year} {2012})}\BibitemShut {NoStop}%
\bibitem [{\citenamefont {Burgo}\ \emph {et~al.}(2012)\citenamefont {Burgo},
  \citenamefont {Ducati}, \citenamefont {Francisco}, \citenamefont
  {Clinckspoor}, \citenamefont {Galembeck},\ and\ \citenamefont
  {Galembeck}}]{burgo_triboelectricity_2012}%
  \BibitemOpen
  \bibfield  {author} {\bibinfo {author} {\bibfnamefont {T.~A.~L.}\
  \bibnamefont {Burgo}}, \bibinfo {author} {\bibfnamefont {T.~R.~D.}\
  \bibnamefont {Ducati}}, \bibinfo {author} {\bibfnamefont {K.~R.}\
  \bibnamefont {Francisco}}, \bibinfo {author} {\bibfnamefont {K.~J.}\
  \bibnamefont {Clinckspoor}}, \bibinfo {author} {\bibfnamefont
  {F.}~\bibnamefont {Galembeck}},\ and\ \bibinfo {author} {\bibfnamefont
  {S.~E.}\ \bibnamefont {Galembeck}},\ }\bibfield  {title} {\bibinfo {title}
  {Triboelectricity: {Macroscopic} {Charge} {Patterns} {Formed} by
  {Self}-{Arraying} {Ions} on {Polymer} {Surfaces}},\ }\href
  {https://doi.org/10.1021/la301228j} {\bibfield  {journal} {\bibinfo
  {journal} {Langmuir}\ }\textbf {\bibinfo {volume} {28}},\ \bibinfo {pages}
  {7407} (\bibinfo {year} {2012})}\BibitemShut {NoStop}%
\bibitem [{\citenamefont {Balestrin}\ \emph {et~al.}(2014)\citenamefont
  {Balestrin}, \citenamefont {Duque}, \citenamefont {Silva},\ and\
  \citenamefont {Galembeck}}]{balestrin_triboelectricity_2014}%
  \BibitemOpen
  \bibfield  {author} {\bibinfo {author} {\bibfnamefont {L.~B. d.~S.}\
  \bibnamefont {Balestrin}}, \bibinfo {author} {\bibfnamefont {D.~D.}\
  \bibnamefont {Duque}}, \bibinfo {author} {\bibfnamefont {D.~S.~d.}\
  \bibnamefont {Silva}},\ and\ \bibinfo {author} {\bibfnamefont
  {F.}~\bibnamefont {Galembeck}},\ }\bibfield  {title} {\bibinfo {title}
  {Triboelectricity in insulating polymers: evidence for a mechanochemical
  mechanism},\ }\href {https://doi.org/10.1039/C3FD00118K} {\bibfield
  {journal} {\bibinfo  {journal} {Faraday Discuss.}\ }\textbf {\bibinfo
  {volume} {170}},\ \bibinfo {pages} {369} (\bibinfo {year}
  {2014})}\BibitemShut {NoStop}%
\bibitem [{\citenamefont {Xie}\ \emph {et~al.}(2013)\citenamefont {Xie},
  \citenamefont {Li}, \citenamefont {Bao},\ and\ \citenamefont
  {Zhou}}]{xie_contact_2013}%
  \BibitemOpen
  \bibfield  {author} {\bibinfo {author} {\bibfnamefont {L.}~\bibnamefont
  {Xie}}, \bibinfo {author} {\bibfnamefont {G.}~\bibnamefont {Li}}, \bibinfo
  {author} {\bibfnamefont {N.}~\bibnamefont {Bao}},\ and\ \bibinfo {author}
  {\bibfnamefont {J.}~\bibnamefont {Zhou}},\ }\bibfield  {title} {\bibinfo
  {title} {Contact electrification by collision of homogenous particles},\
  }\href {https://doi.org/10.1063/1.4804331} {\bibfield  {journal} {\bibinfo
  {journal} {Journal of Applied Physics}\ }\textbf {\bibinfo {volume} {113}},\
  \bibinfo {pages} {184908} (\bibinfo {year} {2013})}\BibitemShut {NoStop}%
\bibitem [{\citenamefont {Yu}\ \emph {et~al.}(2017)\citenamefont {Yu},
  \citenamefont {Mu},\ and\ \citenamefont {Xie}}]{yu_numerical_2017}%
  \BibitemOpen
  \bibfield  {author} {\bibinfo {author} {\bibfnamefont {H.}~\bibnamefont
  {Yu}}, \bibinfo {author} {\bibfnamefont {L.}~\bibnamefont {Mu}},\ and\
  \bibinfo {author} {\bibfnamefont {L.}~\bibnamefont {Xie}},\ }\bibfield
  {title} {\bibinfo {title} {Numerical simulation of particle size effects on
  contact electrification in granular systems},\ }\href
  {https://doi.org/10.1016/j.elstat.2017.10.001} {\bibfield  {journal}
  {\bibinfo  {journal} {J. Electrostat.}\ }\textbf {\bibinfo {volume} {90}},\
  \bibinfo {pages} {113} (\bibinfo {year} {2017})}\BibitemShut {NoStop}%
\bibitem [{\citenamefont {Harris}\ \emph {et~al.}(2019)\citenamefont {Harris},
  \citenamefont {Lim},\ and\ \citenamefont {Jaeger}}]{harris_temperature_2019}%
  \BibitemOpen
  \bibfield  {author} {\bibinfo {author} {\bibfnamefont {I.~A.}\ \bibnamefont
  {Harris}}, \bibinfo {author} {\bibfnamefont {M.~X.}\ \bibnamefont {Lim}},\
  and\ \bibinfo {author} {\bibfnamefont {H.~M.}\ \bibnamefont {Jaeger}},\
  }\bibfield  {title} {\bibinfo {title} {Temperature dependence of nylon and
  {PTFE} triboelectrification},\ }\href
  {https://doi.org/10.1103/PhysRevMaterials.3.085603} {\bibfield  {journal}
  {\bibinfo  {journal} {Phys. Rev. Mater.}\ }\textbf {\bibinfo {volume} {3}},\
  \bibinfo {pages} {085603} (\bibinfo {year} {2019})}\BibitemShut {NoStop}%
\bibitem [{\citenamefont {Grosjean}\ \emph {et~al.}(2020)\citenamefont
  {Grosjean}, \citenamefont {Wald}, \citenamefont {Sobarzo},\ and\
  \citenamefont {Waitukaitis}}]{grosjean_quantitatively_2020}%
  \BibitemOpen
  \bibfield  {author} {\bibinfo {author} {\bibfnamefont {G.}~\bibnamefont
  {Grosjean}}, \bibinfo {author} {\bibfnamefont {S.}~\bibnamefont {Wald}},
  \bibinfo {author} {\bibfnamefont {J.~C.}\ \bibnamefont {Sobarzo}},\ and\
  \bibinfo {author} {\bibfnamefont {S.}~\bibnamefont {Waitukaitis}},\
  }\bibfield  {title} {\bibinfo {title} {Quantitatively consistent
  scale-spanning model for same-material tribocharging},\ }\href
  {https://doi.org/10.1103/PhysRevMaterials.4.082602} {\bibfield  {journal}
  {\bibinfo  {journal} {Phys. Rev. Materials}\ }\textbf {\bibinfo {volume}
  {4}},\ \bibinfo {pages} {082602(R)} (\bibinfo {year} {2020})}\BibitemShut
  {NoStop}%
\bibitem [{\citenamefont {Hull}(1949)}]{hull_method_1949}%
  \BibitemOpen
  \bibfield  {author} {\bibinfo {author} {\bibfnamefont {H.~H.}\ \bibnamefont
  {Hull}},\ }\bibfield  {title} {\bibinfo {title} {A {Method} for {Studying}
  the {Distribution} and {Sign} of {Static} {Charges} on {Solid} {Materials}},\
  }\href {https://doi.org/10.1063/1.1698290} {\bibfield  {journal} {\bibinfo
  {journal} {Journal of Applied Physics}\ }\textbf {\bibinfo {volume} {20}},\
  \bibinfo {pages} {1157} (\bibinfo {year} {1949})}\BibitemShut {NoStop}%
\bibitem [{\citenamefont {Terris}\ \emph {et~al.}(1989)\citenamefont {Terris},
  \citenamefont {Stern}, \citenamefont {Rugar},\ and\ \citenamefont
  {Mamin}}]{terris_contact_1989}%
  \BibitemOpen
  \bibfield  {author} {\bibinfo {author} {\bibfnamefont {B.~D.}\ \bibnamefont
  {Terris}}, \bibinfo {author} {\bibfnamefont {J.~E.}\ \bibnamefont {Stern}},
  \bibinfo {author} {\bibfnamefont {D.}~\bibnamefont {Rugar}},\ and\ \bibinfo
  {author} {\bibfnamefont {H.~J.}\ \bibnamefont {Mamin}},\ }\bibfield  {title}
  {\bibinfo {title} {Contact electrification using force microscopy},\ }\href
  {https://doi.org/10.1103/PhysRevLett.63.2669} {\bibfield  {journal} {\bibinfo
   {journal} {Phys. Rev. Lett.}\ }\textbf {\bibinfo {volume} {63}},\ \bibinfo
  {pages} {2669} (\bibinfo {year} {1989})}\BibitemShut {NoStop}%
\bibitem [{\citenamefont {Shinbrot}\ \emph {et~al.}(2008)\citenamefont
  {Shinbrot}, \citenamefont {Komatsu},\ and\ \citenamefont
  {Zhao}}]{shinbrot_spontaneous_2008}%
  \BibitemOpen
  \bibfield  {author} {\bibinfo {author} {\bibfnamefont {T.}~\bibnamefont
  {Shinbrot}}, \bibinfo {author} {\bibfnamefont {T.~S.}\ \bibnamefont
  {Komatsu}},\ and\ \bibinfo {author} {\bibfnamefont {Q.}~\bibnamefont
  {Zhao}},\ }\bibfield  {title} {\bibinfo {title} {Spontaneous tribocharging of
  similar materials},\ }\href {https://doi.org/10.1209/0295-5075/83/24004}
  {\bibfield  {journal} {\bibinfo  {journal} {EPL}\ }\textbf {\bibinfo {volume}
  {83}},\ \bibinfo {pages} {24004} (\bibinfo {year} {2008})}\BibitemShut
  {NoStop}%
\bibitem [{\citenamefont {Barnes}\ and\ \citenamefont
  {Dinsmore}(2016)}]{barnes_heterogeneity_2016}%
  \BibitemOpen
  \bibfield  {author} {\bibinfo {author} {\bibfnamefont {A.~M.}\ \bibnamefont
  {Barnes}}\ and\ \bibinfo {author} {\bibfnamefont {A.~D.}\ \bibnamefont
  {Dinsmore}},\ }\bibfield  {title} {\bibinfo {title} {Heterogeneity of surface
  potential in contact electrification under ambient conditions: {A} comparison
  of pre- and post-contact states},\ }\href
  {https://doi.org/10.1016/j.elstat.2016.04.002} {\bibfield  {journal}
  {\bibinfo  {journal} {Journal of Electrostatics}\ }\textbf {\bibinfo {volume}
  {81}},\ \bibinfo {pages} {76} (\bibinfo {year} {2016})}\BibitemShut {NoStop}%
\bibitem [{\citenamefont {Pertl}\ \emph {et~al.}(2022)\citenamefont {Pertl},
  \citenamefont {Sobarzo}, \citenamefont {Shafeek}, \citenamefont {Cramer},\
  and\ \citenamefont {Waitukaitis}}]{pertl_quantifying_2022}%
  \BibitemOpen
  \bibfield  {author} {\bibinfo {author} {\bibfnamefont {F.}~\bibnamefont
  {Pertl}}, \bibinfo {author} {\bibfnamefont {J.~C.}\ \bibnamefont {Sobarzo}},
  \bibinfo {author} {\bibfnamefont {L.}~\bibnamefont {Shafeek}}, \bibinfo
  {author} {\bibfnamefont {T.}~\bibnamefont {Cramer}},\ and\ \bibinfo {author}
  {\bibfnamefont {S.}~\bibnamefont {Waitukaitis}},\ }\bibfield  {title}
  {\bibinfo {title} {Quantifying nanoscale charge density features of
  contact-charged surfaces with an {FEM}/{KPFM}-hybrid approach},\ }\href
  {https://doi.org/10.1103/PhysRevMaterials.6.125605} {\bibfield  {journal}
  {\bibinfo  {journal} {Phys. Rev. Mater.}\ }\textbf {\bibinfo {volume} {6}},\
  \bibinfo {pages} {125605} (\bibinfo {year} {2022})}\BibitemShut {NoStop}%
\bibitem [{\citenamefont {Sobolev}\ \emph {et~al.}(2022)\citenamefont
  {Sobolev}, \citenamefont {Adamkiewicz}, \citenamefont {Siek},\ and\
  \citenamefont {Grzybowski}}]{sobolev_charge_2022}%
  \BibitemOpen
  \bibfield  {author} {\bibinfo {author} {\bibfnamefont {Y.~I.}\ \bibnamefont
  {Sobolev}}, \bibinfo {author} {\bibfnamefont {W.}~\bibnamefont
  {Adamkiewicz}}, \bibinfo {author} {\bibfnamefont {M.}~\bibnamefont {Siek}},\
  and\ \bibinfo {author} {\bibfnamefont {B.~A.}\ \bibnamefont {Grzybowski}},\
  }\bibfield  {title} {\bibinfo {title} {Charge mosaics on contact-electrified
  dielectrics result from polarity-inverting discharges},\ }\href
  {https://doi.org/10.1038/s41567-022-01714-9} {\bibfield  {journal} {\bibinfo
  {journal} {Nat. Phys.}\ }\textbf {\bibinfo {volume} {18}},\ \bibinfo {pages}
  {1347} (\bibinfo {year} {2022})}\BibitemShut {NoStop}%
\bibitem [{\citenamefont {Grosjean}\ and\ \citenamefont
  {Waitukaitis}(2023)}]{grosjean_single-collision_2023}%
  \BibitemOpen
  \bibfield  {author} {\bibinfo {author} {\bibfnamefont {G.}~\bibnamefont
  {Grosjean}}\ and\ \bibinfo {author} {\bibfnamefont {S.}~\bibnamefont
  {Waitukaitis}},\ }\bibfield  {title} {\bibinfo {title} {Single-{Collision}
  {Statistics} {Reveal} a {Global} {Mechanism} {Driven} by {Sample} {History}
  for {Contact} {Electrification} in {Granular} {Media}},\ }\href
  {https://doi.org/10.1103/PhysRevLett.130.098202} {\bibfield  {journal}
  {\bibinfo  {journal} {Phys. Rev. Lett.}\ }\textbf {\bibinfo {volume} {130}},\
  \bibinfo {pages} {098202} (\bibinfo {year} {2023})}\BibitemShut {NoStop}%
\bibitem [{\citenamefont {Shaw}\ \emph {et~al.}(1928)\citenamefont {Shaw},
  \citenamefont {Jex},\ and\ \citenamefont
  {Hardy}}]{shaw_tribo-electricity_1928}%
  \BibitemOpen
  \bibfield  {author} {\bibinfo {author} {\bibfnamefont {P.~E.}\ \bibnamefont
  {Shaw}}, \bibinfo {author} {\bibfnamefont {C.~S.}\ \bibnamefont {Jex}},\ and\
  \bibinfo {author} {\bibfnamefont {W.~B.}\ \bibnamefont {Hardy}},\ }\bibfield
  {title} {\bibinfo {title} {Tribo-electricity and friction. {II}.—{Glass}
  and solid elements},\ }\href {https://doi.org/10.1098/rspa.1928.0037}
  {\bibfield  {journal} {\bibinfo  {journal} {Proc. R. Soc. A}\ }\textbf
  {\bibinfo {volume} {118}},\ \bibinfo {pages} {97} (\bibinfo {year}
  {1928})}\BibitemShut {NoStop}%
\bibitem [{\citenamefont {Lowell}\ and\ \citenamefont
  {Truscott}(1986{\natexlab{a}})}]{lowell_triboelectrification_1986-1}%
  \BibitemOpen
  \bibfield  {author} {\bibinfo {author} {\bibfnamefont {J.}~\bibnamefont
  {Lowell}}\ and\ \bibinfo {author} {\bibfnamefont {W.~S.}\ \bibnamefont
  {Truscott}},\ }\bibfield  {title} {\bibinfo {title} {Triboelectrification of
  identical insulators. {I}. {An} experimental investigation},\ }\href
  {https://doi.org/10.1088/0022-3727/19/7/017} {\bibfield  {journal} {\bibinfo
  {journal} {J. Phys. D: Appl. Phys.}\ }\textbf {\bibinfo {volume} {19}},\
  \bibinfo {pages} {1273} (\bibinfo {year} {1986}{\natexlab{a}})}\BibitemShut
  {NoStop}%
\bibitem [{\citenamefont {Lowell}\ and\ \citenamefont
  {Truscott}(1986{\natexlab{b}})}]{lowell_triboelectrification_1986}%
  \BibitemOpen
  \bibfield  {author} {\bibinfo {author} {\bibfnamefont {J.}~\bibnamefont
  {Lowell}}\ and\ \bibinfo {author} {\bibfnamefont {W.~S.}\ \bibnamefont
  {Truscott}},\ }\bibfield  {title} {\bibinfo {title} {Triboelectrification of
  identical insulators. {II}. {Theory} and further experiments},\ }\href
  {https://doi.org/10.1088/0022-3727/19/7/018} {\bibfield  {journal} {\bibinfo
  {journal} {J. Phys. D: Appl. Phys.}\ }\textbf {\bibinfo {volume} {19}},\
  \bibinfo {pages} {1281} (\bibinfo {year} {1986}{\natexlab{b}})}\BibitemShut
  {NoStop}%
\bibitem [{\citenamefont {Grzybowski}\ \emph {et~al.}(2003)\citenamefont
  {Grzybowski}, \citenamefont {Wiles},\ and\ \citenamefont
  {Whitesides}}]{grzybowski_dynamic_2003}%
  \BibitemOpen
  \bibfield  {author} {\bibinfo {author} {\bibfnamefont {B.~A.}\ \bibnamefont
  {Grzybowski}}, \bibinfo {author} {\bibfnamefont {J.~A.}\ \bibnamefont
  {Wiles}},\ and\ \bibinfo {author} {\bibfnamefont {G.~M.}\ \bibnamefont
  {Whitesides}},\ }\bibfield  {title} {\bibinfo {title} {Dynamic
  {Self}-{Assembly} of {Rings} of {Charged} {Metallic} {Spheres}},\ }\href
  {https://doi.org/10.1103/PhysRevLett.90.083903} {\bibfield  {journal}
  {\bibinfo  {journal} {Phys. Rev. Lett.}\ }\textbf {\bibinfo {volume} {90}},\
  \bibinfo {pages} {083903} (\bibinfo {year} {2003})}\BibitemShut {NoStop}%
\bibitem [{\citenamefont {Wiles}\ \emph {et~al.}(2004)\citenamefont {Wiles},
  \citenamefont {Fialkowski}, \citenamefont {Radowski}, \citenamefont
  {Whitesides},\ and\ \citenamefont {Grzybowski}}]{wiles_effects_2004}%
  \BibitemOpen
  \bibfield  {author} {\bibinfo {author} {\bibfnamefont {J.~A.}\ \bibnamefont
  {Wiles}}, \bibinfo {author} {\bibfnamefont {M.}~\bibnamefont {Fialkowski}},
  \bibinfo {author} {\bibfnamefont {M.~R.}\ \bibnamefont {Radowski}}, \bibinfo
  {author} {\bibfnamefont {G.~M.}\ \bibnamefont {Whitesides}},\ and\ \bibinfo
  {author} {\bibfnamefont {B.~A.}\ \bibnamefont {Grzybowski}},\ }\bibfield
  {title} {\bibinfo {title} {Effects of {Surface} {Modification} and {Moisture}
  on the {Rates} of {Charge} {Transfer} between {Metals} and {Organic}
  {Materials}},\ }\href {https://doi.org/10.1021/jp0457904} {\bibfield
  {journal} {\bibinfo  {journal} {J. Phys. Chem. B}\ }\textbf {\bibinfo
  {volume} {108}},\ \bibinfo {pages} {20296} (\bibinfo {year}
  {2004})}\BibitemShut {NoStop}%
\bibitem [{\citenamefont {Thomas~III}\ \emph {et~al.}(2008)\citenamefont
  {Thomas~III}, \citenamefont {Vella}, \citenamefont {Kaufman},\ and\
  \citenamefont {Whitesides}}]{thomas_iii_patterns_2008}%
  \BibitemOpen
  \bibfield  {author} {\bibinfo {author} {\bibfnamefont {S.~W.}\ \bibnamefont
  {Thomas~III}}, \bibinfo {author} {\bibfnamefont {S.~J.}\ \bibnamefont
  {Vella}}, \bibinfo {author} {\bibfnamefont {G.~K.}\ \bibnamefont {Kaufman}},\
  and\ \bibinfo {author} {\bibfnamefont {G.~M.}\ \bibnamefont {Whitesides}},\
  }\bibfield  {title} {\bibinfo {title} {Patterns of {Electrostatic} {Charge}
  and {Discharge} in {Contact} {Electrification}},\ }\href
  {https://doi.org/10.1002/ange.200802062} {\bibfield  {journal} {\bibinfo
  {journal} {Angewandte Chemie}\ }\textbf {\bibinfo {volume} {120}},\ \bibinfo
  {pages} {6756} (\bibinfo {year} {2008})}\BibitemShut {NoStop}%
\end{thebibliography}%

\end{document}